\documentstyle[aps,epsf,floats]{revtex}
\begin{document}

\twocolumn[\hsize\textwidth\columnwidth\hsize\csname@twocolumnfalse%
\endcsname

\title{Aspect-Ratio Scaling and The Stiffness 
Exponent $\theta$ for Ising Spin Glasses}
\author{A. C. Carter, A. J. Bray, M. A. Moore}
\address{Department of Physics \& Astronomy, The University of
Manchester, Manchester M13 9PL, UK}

\date{\today}

\maketitle

\begin{abstract}
We introduce the technique of  {\em aspect-ratio scaling} to study the
scale-dependence of interfacial energies in Ising spin glasses, and we
show how one  can use it to determine  the stiffness exponent $\theta$
in  a   clean  way,   with  results  that   are  independent   of  the
domain-wall-forcing  boundary  conditions imposed  on  the system.  In
space dimension  $d=2$ we obtain  $\theta = -0.282(3)$ for  a Gaussian
distribution of exchange interactions.

\end{abstract}
\vspace{0.5cm} ]

The determination  of stiffness exponents  in spin glasses has  a long
history,  going back  to the  early  1980s \cite{Banavar,McMillan,BM}.
Loosely  speaking,  the  stiffness   exponent  $\theta$  of  an  Ising
spin-glass is defined by the  statement that the energy, $E_{int}$, of
an interface  between ground  states scales with  length scale  $L$ as
$E_{int} \sim L^\theta$.  The  looseness in this definition comes from
the vagueness surrounding the phrase ``length scale $L$''.  Since such
interfaces  are fractals  \cite{FH86,Chaos}, the  length $L$  does not
refer to any  measure on the interface itself, but  rather to the size
of  the region  in  which the  interface  is confined.  Traditionally,
square  or (hyper)cubic  regions of  side  $L$ are  employed, with  an
interface  imposed by a  suitable change  of boundary  condition.  The
problem with this approach is that the results often seem to depend on
the choice of boundary conditions.  Here we explain why this is so and
introduce the technique  of {\em aspect-ratio scaling} as  a method of
obtaining  a  well-defined  value  for $\theta$,  independent  of  the
boundary conditions imposed.

First  we briefly  review some  of the  different  boundary conditions
which have been  used. For simplicity we discuss  only space dimension
$d=2$,  but generalization  to $d>2$  is  trivial.  In  each case  the
boundary  condition  in  the  $y$-direction  is  periodic,  while  the
domain-wall-forcing   boundary   conditions   are   imposed   in   the
$x$-direction.

(i) {\em  Periodic-Antiperiodic} ({\bf P-AP}):  Ground state energies,
$E_P$  and  $E_{AP}$, are  determined  for  periodic and  antiperiodic
boundary  conditions   respectively  (the  latter   often  defined  by
reversing the signs of one column of bonds). The lower-energy state is
obtained for  ${\bf P}$ or  ${\bf AP}$ boundary conditions  with equal
probability,  and  the  higher-energy  state contains  a  domain  wall
relative to the  lower.  The interface energy is  therefore defined as
$E_{int} = |E_{P}-E_{AP}|$.

(ii)  {\em  Free-Antifree}  ({\bf  F-AF}): The  ground  state  energy,
$E_{F}$, and  spin configuration are obtained with  free boundaries in
the $x$-direction. The  spins at one end are held  fixed, those at the
other  end  flipped,  and  the  new  ground  state  energy,  $E_{AF}$,
obtained. In this case $E_{int} = E_{AF} - E_{F}$, and $E_{int} > 0$. 

(iii) {\em Random-Antirandom} ({\bf R-AR}): The spins at both ends are
clamped  in  random   configurations,  and  the  ground-state  energy,
$E_{R}$, is  obtained. The spins at  one end are held  fixed, those at
the other end  flipped, and the new ground-state  energy, $E_{AR}$, is
found. Now $E_{int} = |E_{R}-E_{AR}|$.

We   now   discuss  the   application   of   these   methods  to   the
nearest-neighbor Ising   spin-glass  model   with
Hamiltonian $H =  -\sum_{\langle ij \rangle}J_{ij}S_iS_j$, restricting
our  attention initially to  a Gaussian  distribution of  the exchange
interactions, $J_{ij}$, in dimension $d=2$.  The {\bf P-AP} method has
traditionally been  the most popular.  The exponent  $\theta$ has been
measured for $d=2$ \cite{McMillan,Rieger}, 3 \cite{McMillan,Hartmann3}
and 4 \cite{Hartmann4}.  For $d=2$ the result $\theta = -0.281(2)$ was
obtained  from square  systems of  size $L  \le 30$  \cite{Rieger}.  A
recent  result by  Hartmann and  Young (HY)  \cite{HY} on  much larger
square  systems  ($L \le  480$),  but  using  free boundaries  in  the
$y$-direction, is consistent with this: $\theta = -0.282(2)$.

{\bf R-AR}  boundary conditions were used  in early work by  two of us
\cite{BM} to  obtain $\theta = -0.291(2)$  for $L \le  12$. We believe
(as will be  discussed further below) that the  small discrepancy with
references  \cite{Rieger,HY}  is  due  to  the small  range  of  sizes
available in the earlier study.

Results for {\bf F-AF} boundary  conditions differ from the {\bf P-AP}
results by a  somewhat larger amount, with $\theta  = -0.20$ found for
$L \le 24$  \cite{Matsubara}, and $\theta = -0.266(2)$  obtained by HY
using $L \le 320$ (but with free boundaries in the $y$-direction).  HY
conjecture that  the exponent is actually independent  of the boundary
conditions, but that larger sizes  ($L \gg 320$!)  are needed to reach
the asymptotic regime for {\bf F-AF} boundary conditions.

The questions raised  by these results are (i)  Are the results really
boundary-condition-independent? If  not, what does  $\theta$ mean, and
if  $\theta$  is  not  well-defined,  what  are  we  to  make  of  the
conventional  result   $\xi  \sim  T^{1/\theta}$   \cite{BM}  for  the
divergence of the  correlation length as $T \to  0$?  (ii) If $\theta$
{\em does}  have a well-defined  meaning, independent of  the boundary
conditions,  is there an  efficient method  to obtain  its asymptotic,
boundary-condition-independent value?

To  answer  these  questions  we  introduce  here  the  idea  of  {\em
aspect-ratio  scaling} (ARS).   Using this  approach we  obtain strong
evidence for a unique $\theta$,  and we determine its value as $\theta
= -0.282(3)$,  consistent with values quoted above  from studies using
{\bf P-AP} boundary  conditions.  For a given a  number of spins, this
approach  apparently converges  much  more rapidly  than using  square
samples.

Consider a system  of length $L$ and width  $M$ \cite{Note1}, where we
will usually  take $L \ge  M$. The very natural  assumption underlying
ARS is  that the mean  interfacial energy (averaged over  samples) has
the asymptotic form (for $L$ and $M$ both large)
\begin{equation}
\langle E_{int} \rangle = M^\theta\, F\left(\frac{L}{M}\right)
\label{ARS}
\end{equation}
where $L/M \equiv R$ is the  aspect ratio of the samples. Now consider
the limit  $R \to  \infty$. In  this limit the  system behaves  like a
$d=1$  system,  for  which  one  can  show  that  \cite{BM,Heidelberg}
$\langle  E_{int} \rangle \sim  1/L$. Imposing  this limiting  form on
(\ref{ARS}) requires $F(x) \sim 1/x$ for $x \to \infty$, and gives
\begin{equation}
\langle E_{int} \rangle \sim \frac{M^{1+\theta}}{L},\ \ \ L \gg M\ .
\label{ARS1}
\end{equation}

It is also of interest to consider the limit $M \gg L$. In this limit,
sections of the interface whose spatial extent is much larger than $M$
are essentially independent,  so we expect (for an  $M^{d-1} \times L$
system) the $M$-dependence  $\langle E_{int} \rangle \sim M^{(d-1)/2}$
for {\bf P-AP}  or {\bf R-AR} boundary conditions,  since the energies
of different parts of the interface add with random signs, i.e.\ $F(x)
\sim x^{\theta -(d-1)/2}$ for $x \to 0$ in (\ref{ARS}), to give
\begin{equation}
\langle                                              E_{int}\rangle\sim
L^\theta\,\left(\frac{M}{L}\right)^{(d-1)/2}, \  M \gg L\  \ ({\mathrm
{\bf P-AP}, {\bf R-AR}}),
\label{ARS2}
\end{equation}
while for {\bf  F-AF} boundary conditions they will  add with the same
sign to give
\begin{equation} 
\langle E_{int}\rangle\sim L^\theta\,\left(\frac{M}{L}\right)^{d-1}, \
\ M \gg L\ \ ({\mathrm {\bf F-AF}}).
\label{ARS3}
\end{equation}
Since   the  asymptotic  forms   (\ref{ARS2})  and   (\ref{ARS3})  are
different, it follows that  the scaling function $F(x)$ in (\ref{ARS})
will depend on  the boundary conditions for general  $x$. However, the
limiting  large-$x$ form, which  leads to  (\ref{ARS1}), will  be {\em
independent} of the boundary conditions

If the limit $L  \gg M$ can be achieved in practice,  so that the form
(\ref{ARS1}) holds,  the exponent $\theta$  can be extracted  from the
$M$-dependence,  and  is  transparently  independent of  the  boundary
conditions, i.e.\ if we define
\begin{equation}
G(M) = \lim_{L \to \infty} L\langle E_{int} \rangle\ ,
\end{equation}
then $G(M) \to AM^{1+\theta}$ for $M \to \infty$ ($A=$ const.), giving
\begin{equation}
\theta = \lim_{M \to \infty} \frac{d\ln G}{d\ln M} - 1\ ,
\end{equation}
which is clearly independent of the boundary conditions imposed in the
$x$-direction since the limit $L \to \infty$ is taken before the limit
$M \to \infty$.

In practice we find it convenient to study a broad range of $R$ rather
than  just the regime  $R \gg  1$, though  the exponent  $\theta$ will
ultimately be  obtained by  extrapolation to the  $R =  \infty$ limit.
Figure 1 shows a log-log plot of $L\langle E_{int}\rangle$ against $M$
for various  fixed aspect  ratios $R$  ($1 \le R  \le 32$),  with {\bf
R-AR} boundary conditions. The  ground state energies were obtained by
using exact transfer matrix calculations,  with $2 \le M \le 12$. Each
point is an average of $10^5$ samples.  If ARS works perfectly for all
$L$  and  $M$,  the  lines  corresponding to  different  $R$  will  be
parallel, with slope $1+\theta$.

\begin{figure}[!htb]
\begin{center}
\epsfxsize=8.5cm \epsfbox{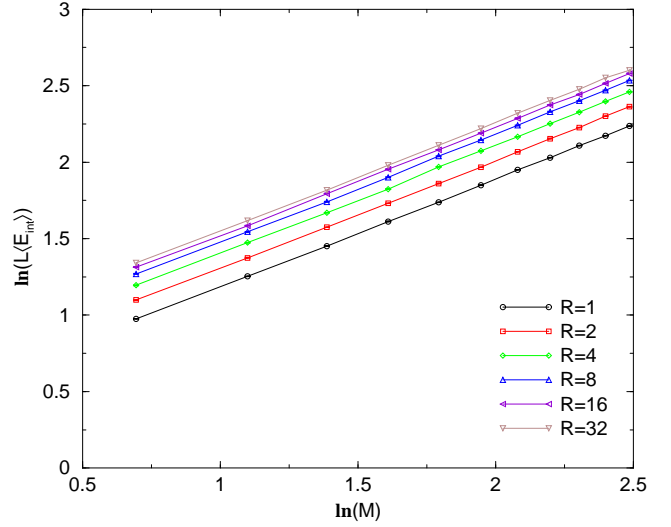}
\caption{Variation  of $\langle E_{int}  \rangle $  with width  $M$ at
fixed $R$ for {\bf R-AR} boundary conditions.
\label{fig:cac}}
\end{center}
\end{figure}

The  lines  are indeed  almost  straight  and  parallel.  The  slopes,
$\theta_R$ ($R=$ ``random'') are presented for different aspect ratios
in Table 1. The $R=1$  result, corresponding to squares, is consistent
with that  obtained earlier using the same  method \cite{BM}. However,
there  is  a  very  slow  decrease  of  the  effective  exponent  with
increasing  $R$. We have  argued that  it is  sensible to  extract the
exponent from the large-$R$ limit,  since in this limit any errors due
to finite-size corrections in  the $x$-direction are eliminated.  Note
that when  the data are truly  in the regime  $R \gg 1$, the  lines in
Figure 1  will fall  on top of  each other.   While they appear  to be
approaching a limit, they have not yet reached the limit at $R=32$.

\begin{table}[!htb]
\begin{tabular}{ccc}
R&$\theta_R$&$\theta_F$\\\hline                 1&-0.289(2)&-0.153(2)\\
2&-0.286(2)&-0.215(2)\\                         4&-0.285(2)&-0.249(2)\\
8&-0.283(2)&-0.265(2)\\                        16&-0.285(2)&-0.273(2)\\
32&-0.283(2)&-0.274(3)\\
\end{tabular}
\caption{Effective stiffness  exponent $\theta$  as a function  of the
aspect ratio R for {\bf R-AR} ($\theta_R$) and {\bf F-AF} ($\theta_F$)
boundary conditions,  obtained from gradients of the  lines in figures
\ref{fig:cac} and \ref{fig:1a}. \label{tab:cac}}
\end{table}

The  equivalent  results  for   {\bf  F-AF}  boundary  conditions  are
presented in  Figure 2. In this  case the lines are  clearly {\em not}
parallel, i.e.\ there is a strong dependence of the measured gradients
on $R$. The  lines are also noticeably curved,  especially for smaller
values of $R$, e.g.\ for  $R=1$ the slope (and therefore the effective
value of $\theta$) is decreasing with increasing $M$.

\begin{figure}[!htb]
\begin{center}
\epsfxsize=8.5cm \epsfbox{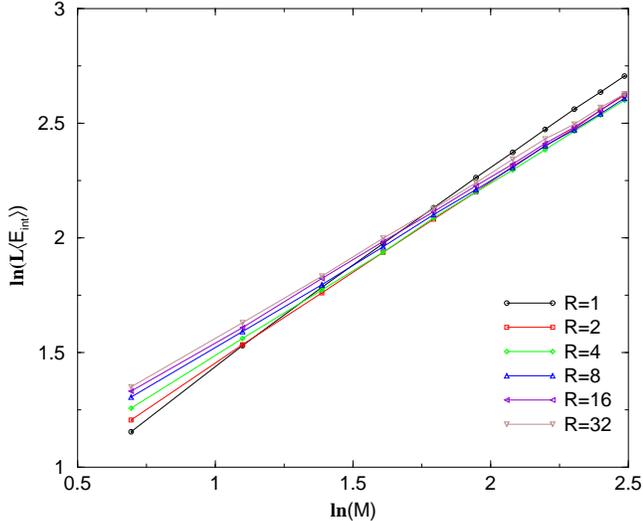}
\caption{Variation  of $\langle  E_{int}  \rangle$ with  width $M$  at
fixed $R$ for {\rm F-AF} boundary conditions.
\label{fig:1a}}
\end{center}
\end{figure}

Table 1 contains the  effective exponents, $\theta_F$ ($F=$ ``free''),
extracted  from Figure  2, for  the  different values  of $R$.   These
exponents  were  obtained  by  fitting  straight lines  to  the  data,
discarding the  smallest value  of $M$ in  each case. Note  the strong
dependence  of  $\theta_F$ on  $R$.   For  $R=1$  (i.e.\ squares)  the
effective $\theta_F$ differs  by almost a factor 2  from the effective
$\theta_R$, but the difference gets  smaller for larger values of $R$,
and we argue that it  approaches zero asymptotically.  The argument is
based on the  following observation.  For large $R$,  we find that the
interface  energy is nearly  always identical,  sample by  sample, for
both boundary conditions.  The differences apparent in Table 1 are due
to a small  fraction of the samples where the  energies differ for the
two  boundary  conditions.   In  fact, inspection  of  the  interfaces
themselves  shows, as one  would expect,  that these  occupy identical
locations  for  the two  boundary  conditions  whenever the  interface
energies are the same. To understand this we note the following facts:
\\ (i)  A detailed study of  the ground states indicates  that, for $R
\gg 1$, the actual spin configurations are independent of the boundary
conditions in the ``interior''  regions away from the boundaries.  The
effect  of changing  the  boundary conditions  is  localized near  the
boundary, propagating  a distance into  the system whose mean  size is
roughly proportional to  $M$ over the range of $M$ ($2  \le M \le 12$)
studied.   \\ (ii)  The width  of the  interfacial region  also scales
roughly as $M$.  \\ One  can understand the latter as follows. Suppose
this width scales as $M^a$.  Then $a<1$ would imply that the interface
is ``smooth''  on large length  scales, inconsistent with it  having a
non-trivial fractal  dimension $d_s >  d-1$. On the other  other hand,
$a>1$  would  imply that  for  $M \gg  1$  the  interface is  strongly
``stretched''  in  the  $x$-direction.    Its  energy  could  then  be
estimated,  using (\ref{ARS2})  with  $M \to  M^a$  and $L  \to M$  as
$E_{int}  \sim  M^{\theta  +  (a-1)(d-1)/2} \gg  M^\theta$,  which  is
inconsistent  \cite{Note2}. We  conclude that  $a=1$. This  means that
there are of order $R = L/M$ independent places in which the interface
can sit.  Any one of these  has an energy of order $M^\theta$, but the
prefactor  is a  random variable  with non-zero  weight at  the origin
\cite{Heidelberg}.  The  smallest of these therefore  scales as $1/R$,
i.e.\ $E_{int} \sim  M^\theta/R$ for $R \gg 1$,  a result identical to
Eq.\ (\ref{ARS1}).  From (i) and (ii) we see that the probability that
the interface enters a region near the boundary where the ground-state
spin configuration  differs for the  two sets of  boundary conditions,
and for which the interace energies also differ, is of order $1/R$ for
large $R$.

\begin{figure}[!htb]
\begin{center}
\epsfxsize=8.5cm \epsfbox{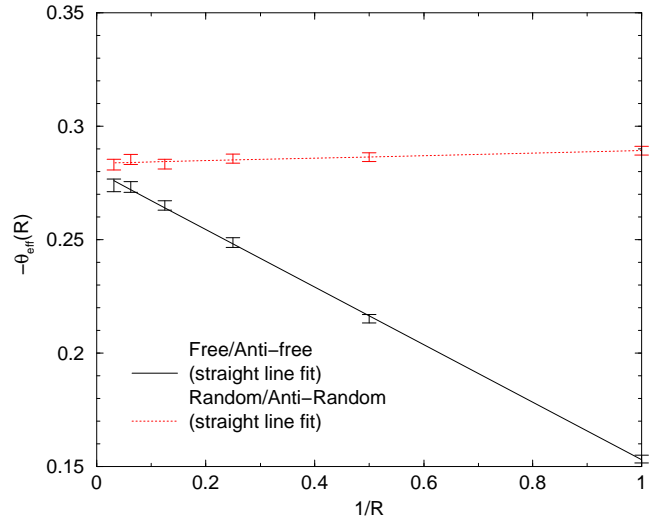}
\caption{Variation  of   $\theta_{\mathrm{eff}}$  with  system  aspect
ratio, $R$,  for {\bf R-AR} (upper  data) and {\bf  F-AF} (lower data)
boundary  conditions. The data  are consistent  with convergence  to a
unique limiting value, $\theta = -0.282(3)$.}
\end{center}
\end{figure}

The  above considerations  suggest  that convergence  to the  limiting
behavior  (\ref{ARS1}) occurs as  $1/R$, so  in Figure  3 we  plot the
effective exponents  listed in Table  1 against $1/R$.   The resulting
plots are compatible  with a linear dependence on  $1/R$, and the data
are  consistent  with  convergence   to  a  unique  value,  $\theta  =
-0.282(3)$, as $R \to \infty$.

It  is remarkable  that, by  exploiting ARS,  studies of  systems with
relatively  small  widths, $2  \le  M \le  12$,  can  give results  of
comparable  precision to  studies  on square  systems  of much  larger
size. This is especially  striking for {\bf F-AF} boundary conditions,
where the effective exponent depends strongly on the aspect ratio $R$.
For  example, Table 1  shows that  already for  $R=2$ the  estimate of
$\theta$ is more accurate than  the result $\theta_F = -0.20$ obtained
for square systems of size  up to 24 \cite{Matsubara}, while for $R=8$
it matches the  estimate $\theta_F = -0.266$ obtained  from squares up
to size 320 \cite{HY}.  Furthermore, the ARS method demonstrates quite
convincingly  that   the  stiffness  exponent   is  boundary-condition
independent, a  result which has  not been confirmed  conclusively for
squares  even on  the  largest systems  studied  \cite{HY}. This  slow
convergence  of  $\theta$ for  {\bf  F-AF}  boundary conditions  using
squares  suggests  that other  methods  of  determining $\theta$,  for
example from the  dependence on system size $L$  of the Parisi overlap
function  $P(q)$  at  $q=0$  (notably  in  $d=3$),  using  $P(0)  \sim
L^{-\theta}$,     as     predicted     by    the     droplet     model
\cite{FH86,Heidelberg},  or creating  droplets by  flipping  a central
spin while  holding the  boundary spins fixed  \cite{Kawashima}, might
also suffer from large finite-size effects. 

We conclude by presenting some  results for $d=3$. The transfer matrix
approach  restricts us to  rather small  widths, $M  \le 4$.  For each
value of $R$ we  obtain an effective stiffness exponent $\theta_{eff}$
by fitting a straight line to  the three points $M=2,3,4$ in a plot of
$\ln(L\langle |E| \rangle)$ against $\ln M$, and defining the slope to
be $1 + \theta_{eff}$. These lines  are found to be quite straight, so
$\theta_{eff}$ can  be readily extracted. It is  plotted against $1/R$
in  Figure  4,  for  both  ${\bf  R-AR}$  and  ${\bf  F-AF}$  boundary
conditions.   Also plotted  are the  equivalent results  for  the case
where {\em free}  boundaries are used in the  directions normal to the
domain-wall forcing  boundary conditions.  The final  character in the
legend  specifies  whether  periodic  ({\bf  P})  or  free  ({\bf  F})
boundaries have been employed  in these directions.  The free boundary
data lie above the corresponding periodic boundary data in each case.
\begin{figure}[!htb]
\begin{center}
\epsfxsize=8.5cm \epsfbox{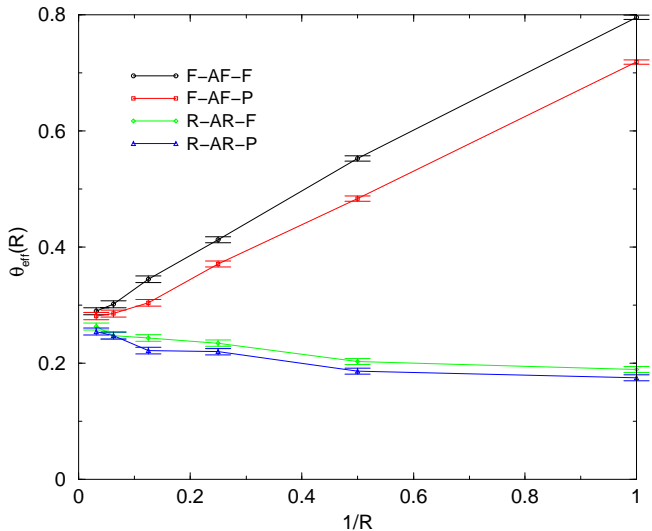}
\caption{Variation  of   $\theta_{\mathrm{eff}}$  with  system  aspect
ratio, $R$, in $d=3$ for {\bf R-AR} (upper  data) and {\bf  F-AF} 
(lower data) boundary  conditions. Periodic and free boundary conditions 
in the $y$ direction are indicated by the final character ${\bf P}$ or 
${\bf F}$.} 
\end{center}
\end{figure}
The dependence of $\theta_{eff}$ on $R$ in this case is quite striking. 
For {\bf R-AR-P} boundary conditions, the $R=1$ result, $\theta_{eff} 
\simeq 0.19$ is in agreement with our earlier results \cite{BM} on 
cubes of side $L=2,3,4$ and those of Hartmann for $L \le 10$ 
\cite{Hartmann3}. For $R \to \infty$, however, results for all boundary 
seem to converge to value $\theta \simeq 0.27$, significantly larger 
than previous estimates. It is surprising, also, that the boundary 
conditions in the transverse direction do not significantly affect the 
large-$R$ limit of $\theta_{eff}$. While the small widths $M$ used prompts 
caution in the interpretation of this result (the same widths, used for 
$d=2$, would give $\theta \simeq -0.32$ instead of $-0.28$), the trends 
with increasing $R$ are quite striking. In particular, the use of 
${\bf F-AF}$ boundary conditions for cubes (i.e.\ $R=1$) leads to a 
very large overestimate of the exponent. 

In summary,  the use of aspect-ratio scaling gives, in $d=2$, results 
comparable in quality to those obtained from square systems of much 
larger size, and independent of the domain-wall-forcing boundary condition. 
It does this by eliminating finite-size corrections in the direction normal 
to the domain wall. For $d=3$ the results suggest that $\theta$ may be 
significantly larger than previous estimates. 

We thank A. P. Young for a useful discussion. This work was stimulated
in  part by  comments from  J. M. Kosterlitz concerning  the possible
influence  of boundary  conditions on  the determination  of stiffness
exponents, and was supported by EPSRC.

\end{document}